\documentclass[twocolumn,showpacs,preprintnumbers,amsmath,amssymb,superscriptaddress]{revtex4}

\usepackage{graphicx,dcolumn,bm}

\newcommand{\nl}{\nonumber \\}
\newcommand{\be}{\begin{equation}}
\newcommand{\ee}{\end{equation}}
\newcommand{\bea}{\begin{eqnarray}}
\newcommand{\eea}{\end{eqnarray}}
\newcommand{\bsube}{\begin{subequations}}
\newcommand{\esube}{\end{subequations}}

\newcommand{\Eq}[1]{Eq.\,(\ref{#1})}
\newcommand{\Eqs}[1]{Eqs.\,(\ref{#1})}
\newcommand{\fref}[1]{Fig.\,\ref{#1}}
\newcommand{\Fig}[1]{Fig.\,\ref{#1}}
\newcommand{\sref}[1]{Sec.\,\ref{#1}}

\newcommand{\rmS}{{\rm S}}
\newcommand{\rmB}{{\rm B}}
\newcommand{\rmL}{{\rm L}}
\newcommand{\rmR}{{\rm R}}

\newcommand{\rmT}{{\rm T}}
\newcommand{\rmc}{{\rm c}}
\newcommand{\rmi}{{\rm i}}
\newcommand{\rmd}{{\rm d}}
\newcommand{\rmq}{{\rm q}}

\newcommand{\NL}{N_{\rm L}}
\newcommand{\NR}{N_{\rm R}}

\newcommand{\alf}{\alpha}
\newcommand{\sgm}{\sigma}
\newcommand{\Omg}{\Omega}
\newcommand{\omg}{\omega}
\newcommand{\Gam}{\Gamma}

\newcommand{\vpl}{\varepsilon}
\newcommand{\epl}{\epsilon}

\newcommand{\GamL}{\Gamma_{\rm L}}
\newcommand{\GamR}{\Gamma_{\rm R}}

\newcommand{\la}{\langle}
\newcommand{\ra}{\rangle}
\newcommand{\Tr}{{\rm Tr}}
\newcommand{\etaL}{\eta_{\rm L}}
\newcommand{\etaR}{\eta_{\rm R}}

\begin{document}

\title{Qubit detection with a T-shaped double quantum dot detector}


\author{JunYan Luo}\email{jyluo@zust.edu.cn}
\affiliation{Department of Physics, Zhejiang University of Science
  and Technology, Hangzhou 310023, China}
\author{HuJun Jiao}
\affiliation{Department of Physics, Shanxi University, Taiyuan,
 Shanxi 030006, China}
\author{Jing Hu}
\affiliation{Department of Physics, Zhejiang University of Science
  and Technology, Hangzhou 310023, China}
\author{Xiao-Ling He}
\affiliation{Department of Physics, Zhejiang University of Science
  and Technology, Hangzhou 310023, China}
\author{XiaoLi Lang}
\affiliation{Department of Physics, Zhejiang University of Science
  and Technology, Hangzhou 310023, China}
\author{Shi-Kuan Wang}
\affiliation{Department of Physics, Hangzhou Dianzi University,
Hangzhou 310018, China}

\date{\today}




 \begin{abstract}
 We propose to continuously monitor a charge qubit by utilizing a T-shaped double
 quantum dot detector, in which the qubit and double dot are arranged in such a unique
 way that the detector turns out to be particularly susceptible to
 the charge states of the qubit.
 Special attention is paid to the regime where acquisition of
 qubit information and backaction upon the measured system exhibit
 nontrivial correlation.
 The intrinsic dynamics of the qubit gives rise to
 dynamical blockade of tunneling events through the detector, resulting
 in a super-Poissonian noise.
 However, such a pronounced enhancement of detector's shot noise does not
 necessarily produce a rising dephasing rate.
 In contrast, an inhibition of dephasing is entailed by the reduction of
 information acquisition in the dynamically blockaded regimes.
 We further reveal the important impact of the charge fluctuations on
 the measurement characteristics.
 Noticeably, under the condition of symmetric junction capacitances
 the noise pedestal of circuit current is completely suppressed,
 leading to a divergent signal-to-noise ratio, and eventually to
 a violation of the Korotkov-Averin bound in quantum measurement.
 Our study offers the possibility for a double dot detector to reach
 the quantum limited effectiveness in a transparent manner.
\end{abstract}


 \maketitle

 \section{\label{thsec1}Introduction}

 Understanding the fundamental physics in quantum measurement
 process is of vital importance for physically implementing
 fast and efficient measurement of a two-state quantum system
 (qubit) \cite{All131}, as well as essential applications in
 quantum information processing \cite{Wis10,Pet052180}.
 So far, a variety of mesoscopic devices have been proposed
 for fast readout of qubit information.
 For instance, a quantum point contact (QPC) has been widely
 investigated, with special attention paid to the nontrivial
 correlation between the QPC and
 qubit \cite{Cle03165324,Pil02200401,Ave05126803,Li05066803,Tau08176805,%
 Luo09385801,Gus09191,Pet10246804,You10186803,Tho12235419,Ubb12612}.
 Alternatively, a single electron transistor (SET) was shown
 to have advantages over QPC in many respects, such as high
 sensitivity, wide circuit bandwidth, and low
 noise \cite{Dev001039,Mak01357,Cle02176804,Moz04018303,Gur05073303,%
 Jia09075320,Luo104904}.
 In particular, single-shot measurement has recently been realized based
 on SET detectors, in which the information of the qubit is uniquely
 determined in simply  one run \cite{Lu03422,Bie06201402,Mor10687,Deh14236801}.

 Historically, quantum mechanical detection was described by
 the projective theory, in which the measurement takes place
 instantaneously.
 In contrast, the essence of the modern theory of quantum
 measurement emphasizes that detector extracts information
 and renders the measured system in a continuous manner.
 The process of information acquisition from the detector
 and how it would alter the remaining uncertainty in the
 system lies at the heart of the measurement dynamics.
 An important figure of merit in continuous measurement
 is the detector ``ideality'' or effectiveness, characterizing
 how close to the quantum limit the detector could operator.
 In an ideal detection, qubit dephasing generated by detector
 backaction is purely associated with the information flow,
 rather than a noisy environment.
 For a less effective detector, qubit dephasing takes place
 more rapidly than the information flow.
 This imposes an important limit on the signal-to-noise ratio
 of the measurement, known as the Korotkov-Averin bound: The
 maximum signal-to-noise ratio the detector can reach is limited at
 4 \cite{Kor01165310,Kor01085312}.
 It has been confirmed in Refs. \cite{Rus03075303,Shn04840}
 and measured in experiment \cite{Ili03097906}.
 Extensions of the Korotkov-Averin bound have also been
 discussed in continuous measurement of coupled
 qubits \cite{Mao04056803,Mao05085320} and
 precession of an individual spin \cite{Bul04040401,Nus03085402}.

\begin{figure*}
\includegraphics*[scale=0.8]{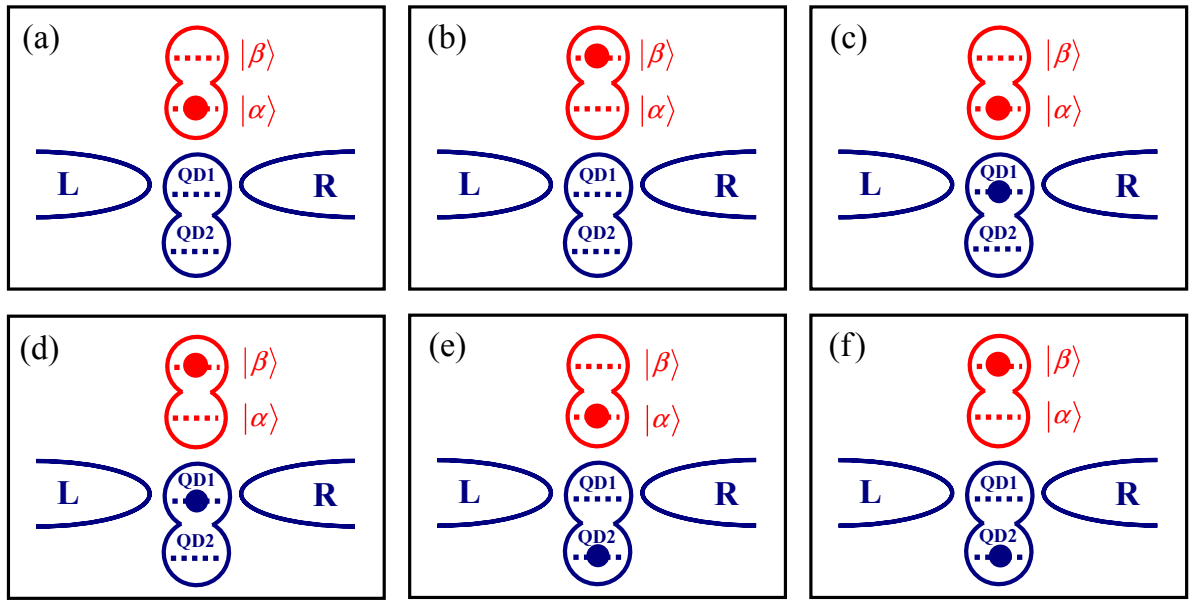}
\caption{\label{Fig1}Schematics of a solid-state charge qubit under
the continuous measurement of a TDQD detector.
 Possible electron configurations of the reduced system (qubit plus TDQD)
 are:  (a) the TDQD is empty, (c) QD1 is occupied, and (e) QD2 is occupied
 while the electron of the qubit resides the logic state $|\alf\ra$.
 Correspondingly, (b), (d) and (f) denote the same states but with
 the electron of the qubit in the state $|\beta\ra$.}
\end{figure*}

 For an SET detector, it usually means a single quantum dot (SQD)
 sandwiched between the source and drain electrodes, in which
 electron transport exhibits quantum coherence within the size
 of the reduced system.
 Yet, the discrete nature of the charge exhibits its inherent
 randomness in the process of transport.
 The involving shot noise and telegraph noise was recently
 proved to be the two sides of the same
 coin \cite{Sch042005,Van044394,Gus06076605,Fuj061634}, and may
 have essential roles to play in the quantum measurement.
 It has been shown that the SQD detector may achieve quantum
 limited measurement under appropriate conditions, where
 qubit dephasing is due purely to the information flow, rather
 than detector's shot noise \cite{Jia07155333,Luo104904}.
 Yet, in order to distinguish  clearly the two currents
 corresponding to the two logical states of the qubit, it poses
 a very challenging condition in measurement, i.e. very low temperature.

 To loosen the tough
 temperature restrictions, a double quantum dot (DQD) SET has recently
 been proposed to continuously monitor a qubit \cite{Jia07155333}.
 The electrostatic interaction between the qubit and DQD leads to
 an energy level mismatch between the two dots, which causes a prominent current
 visibility of the measurement even at a relatively high temperature.
 Unfortunately, its effectiveness turns out to be less than that
 of an ideal detector \cite{Gil06116806}.
 The reason is that the generated dephasing of the qubit stems
 partially from the detector's shot noise, such that the
 information of qubit encoded in the DQD detector's degree
 of freedom cannot be fully deduced from the measured output.
 It is thus appealing to find a detector capable of combining advantages
 of SQD and DQD detectors together, such that it could operate at
 a weakened temperature condition while reaching the maximum
 effectiveness at the same time.

 In this work, we investigate this important issue in the context
 of a T-shaped DQD (TDQD) detector \cite{Luo10083720,Luo1159,Luo13155304,Xue13208},
 where only quantum dot 1 (QD1) is directly tunnel-coupled to the left
 and right electrodes, whereas quantum dot 2 (QD2) is only
 side-coupled to QD1 (see \fref{Fig1}).
 We pay special attention to the essential correlation between
 the qubit and the TDQD detector.
 In particular, the inherent dynamics of the qubit may give rise
 to bunching of tunneling events though the TDQD detector, which
 is manifested as a pronounced super-Poissonian noise in the TDQD
 detector.
 However, such a large noise does not necessarily imply an enhancement
 of the dephasing rate. In contrast, the involving dynamical blockade
 corresponds to a no-measurement regime, where the qubit dephasing is
 actually suppressed.
 An important advantage of the SET detector is that
 the left and right electrodes could monitor the qubit simultaneously,
 such that any noise not shared by two electrodes can be filtered out,
 making it analogous to the measurement setup of twin quantum point
 contacts \cite{Jor05220401}.
 However, the crucial difference is that the currents through the
 left and right junctions of the TDQD detector are intrinsically correlated to
 each other via the charge fluctuations in the TDQD.
 We demonstrate that
 although the signal-to-noise ratio associated with the junction
 noise alone could not approach the quantum limit, the spectrum of
 charge fluctuations in the TDQD results in a complete suppression
 of the the noise pedestal, leading eventually to a divergent
 signal-to-noise ratio and thus a violation of the Korotkov-Averin bound.

 The paper is organized as follows.
 We start in \sref{thsec2} with a description the measurement setup
 and corresponding Hamiltonian for this scenario.
 The particle-number-resolved quantum master equation (QME) to
 the reduced dynamics and measurement characteristics is
 outlined in \sref{thsec3}.
 The influence of qubit dynamics
 on the TDQD detector shot noise is analyzed in \sref{thsec4}, which is then
 followed by the discussion qubit dephasing behavior
 associated with detector's output in \sref{thsec5}.
 \sref{thsec6} is focused on the measurement
 effectiveness of the TDQD detector in terms of the signal-to-noise
 ratio, with special attention paid to the violation of the Korotkov-Averin bound.
 Finally, we conclude in \sref{thsec7}.

\begin{table*}
\caption{\label{tab1}The eigenenergies and corresponding eigenstates
of the reduced system (qubit plus TDQD) for $\epl_\rmq=\epl_\rmT=0$, and
 $\Omg$, $W\ll U$.}
\begin{ruledtabular}
\begin{tabular}{ccc}
 $N_{\rm tot}$ & Eigenenergy & Eigenstate \\ \hline
 1&$E_1=-\Omg$ & $|1\ra=\frac{1}{\sqrt{2}}(|{\rm a}\ra-|{\rm b}\ra)$
 \vspace{0.1cm}\\
 2&$E_2=+\Omg$ & $|2\ra=\frac{1}{\sqrt{2}}(|{\rm a}\ra+|{\rm b}\ra)$
 \vspace{0.1cm}\\
 3& $E_3\simeq 0 $ &$|3\ra=\frac{1}{\sqrt{2}}(|{\rm d}\ra-|{\rm e}\ra)$
 \vspace{0.1cm}\\
 4&$E_4\simeq U$&
 $|4\ra \simeq |{\rm c}\ra$
 \vspace{0.1cm}\\
 5&$E_5\simeq +\frac{1}{\sqrt{2}}(\Omg+W)$ &
 $|5\ra\simeq \frac{1}{2}(|{\rm d}\ra+|{\rm e}\ra+\sqrt{2}|{\rm f}\ra)$
 \vspace{0.1cm}\\
 6&$E_6\simeq -\frac{1}{\sqrt{2}}(\Omg+W)$ &
 \!\!$|6\ra\simeq \frac{1}{2}(|{\rm d}\ra+|{\rm e}\ra-\sqrt{2}|{\rm f}\ra)$
 \\
\end{tabular}
\end{ruledtabular}
\end{table*}

\section{\label{thsec2}Model Description}

 The system under study is  shown schematically in \Fig{Fig1}.
 The charge qubit is represented by an extra electron in a double
 quantum dot.
 Whenever the electron occupies the lower (upper) dot, the qubit
 is said to be in the logic state $|\alf\ra$ ($|\beta\ra$).
 The detector is a TDQD SET, in which QD1 is directly tunnel-coupled
 to the left (L) and right (R) electrodes, whereas QD2 is only
 side-coupled to QD1.
 We assume that each quantum dot has only one level
 involved in transport within the bias window defined
 by the Fermi levels of the left and right electrodes.
 Furthermore, both interdot and intradot charging energies
 are much larger than the Fermi levels such that at most
 one electron can reside on the TDQD.
 The Hilbert space of the TDQD dot is thus reduced to
 $|0\ra$-empty, $|1\ra$ ($|2\ra$)-one electron in QD1 (QD2).
 The qubit is placed in vicinity of QD1, as shown in \Fig{Fig1}.
 Under such a unique arrangement the measured current is expected to be
 particularly susceptible to electron configurations of the qubit.
 It is right this mechanism that can be utilized to
 sensitively acquire the qubit-state information from
 the output of the TDQD detector.

 The entire system Hamiltonian reads
 \be\label{Htot}
 H=H_\rmS+H_\rmB+H'.
 \ee
 The first part denotes the Hamiltonian of the reduced
 system (qubit plus TDQD)
 \bea\label{Hs}
 H_\rmS\!=\!\frac{1}{2}\epl_\rmq \sgm_z+\Omg \sgm_x
 \!+\!\frac{1}{2}\epl_\rmT Q_z\!+\!W Q_x
 \!+\!U |\alf\ra\la\alf|\!\otimes\!|1\ra\la1|,
 \eea
 where we have introduced pseudo-spin operators
 $\sgm_z\equiv|\alf\ra\la\alf|-|\beta\ra\la\beta|$,
 $\sgm_x\equiv|\alf\ra\la\beta|+|\beta\ra\la\alf|$
 for the qubit, and likewise for the TDQD
 $Q_z\equiv|1\ra\la1|-|2\ra\la2|$,
 $Q_x\equiv|1\ra\la2|+|2\ra\la1|$.
 The level detuning and interdot coupling in the qubit (TDQD)
 are $\epl_\rmq$ ($\epl_\rmT$) and $\Omg_\rmq$ ($W$),
 respectively.
 The qubit is placed in close proximity to the QD1, such that the energy
 level of QD1 is very sensitive to the qubit occupations,
 as represented by the last term in \Eq{Hs}.
 There are totally six possible electron configurations of the
 reduced system (qubit plus TDQD), as shown in \Fig{Fig1}(a)-(f).
 Let \{$|{\rm a}\ra$,$\cdots$,$|{\rm f}\ra$\} be the
 states of the reduced system corresponding to charge
 configurations in \Fig{Fig1}(a)-(f).
 The eigenenergies and corresponding eigenstates of the reduced system
 are listed in Table \ref{tab1} for $\epl_\rmq=\epl_\rmT=0$,
 and $\Omg$, $W \ll U$.

 The electrodes are modeled as reservoirs of noninteracting electrons
 \bea
 H_\rmB=\sum_{\ell=\rmL,\rmR}\sum_k\vpl_{\ell k} c_{\ell k}^\dag c_{\ell k},
 \eea
 where $c_{\ell k}^\dag $ ($c_{\ell k}$) stands for the creation
 (annihilation) operator for an electron with momentum $k$ in the 
 left ($\ell$=L) or right ($\ell=$R) electrode.
 The left/right electron reservoir is characterized by the Fermi
 distribution $f_{\rmL/\rmR}(\omg)$.
 The voltage is symmetrically applied, which leads to symmetric
 Fermi levels
 in the left and right electrodes, i.e. $\mu_{\rmL/\rmR}=\pm eV/2$.

 Electron tunneling between the QD1 and electrodes is described by
 \bea\label{Hprm}
 H'=\sum_{\ell,k}(t_{\ell k}c^\dag_{\ell k}|0\ra\la1|+{\rm h.c.})
 \equiv\sum_\ell (f_\ell |0\ra\la1|+{\rm h.c.}),
 \eea
 where $f_\ell\equiv \sum_k t_{\ell k}c^\dag_{\ell k}$.
 The tunnel-coupling strength between electrode $\ell=$\{L,R\} and
 QD1 is given
 by the intrinsic tunneling width
 $\Gam_\ell(\omg)=2\pi \sum_k |t_{\ell k}|^2\delta(\omg-\vpl_{\ell k})$.
 Hereafter, we consider wide band in the electrodes, which
 results in energy independent couplings $\Gam_{\rmL/\rmR}$.
 The total tunneling width  is thus given by $\Gam=\GamL+\GamR$.
 The effects of stochastic electron reservoirs on the measurement
 are characterized by the bath correlation functions
 \begin{subequations}
 \begin{gather}
 C_\ell^{(+)}(t-\tau)= \la f_\ell^\dag(t) f_\ell(\tau)\ra_\rmB,
 \\
 C_\ell^{(-)}(t-\tau)= \la f_\ell(t) f^\dag_\ell(\tau)\ra_\rmB,
 \end{gather}
 \end{subequations}
 where $\la \cdots\ra_\rmB\equiv{\rm tr}_{\rm B}[(\cdots)\rho_{\rm B}]$
 stands for the trace over degrees of freedom of the electron reservoirs,
 with $\rho_{\rm B}$  the local thermal equilibrium state of the electrodes.
 Throughout this work, we set $\hbar=e=1$ for the
 Planck constant and electron charge, unless stated otherwise.

\begin{figure*}
\begin{minipage}[c]{.51\textwidth}
\includegraphics[scale=1.05]{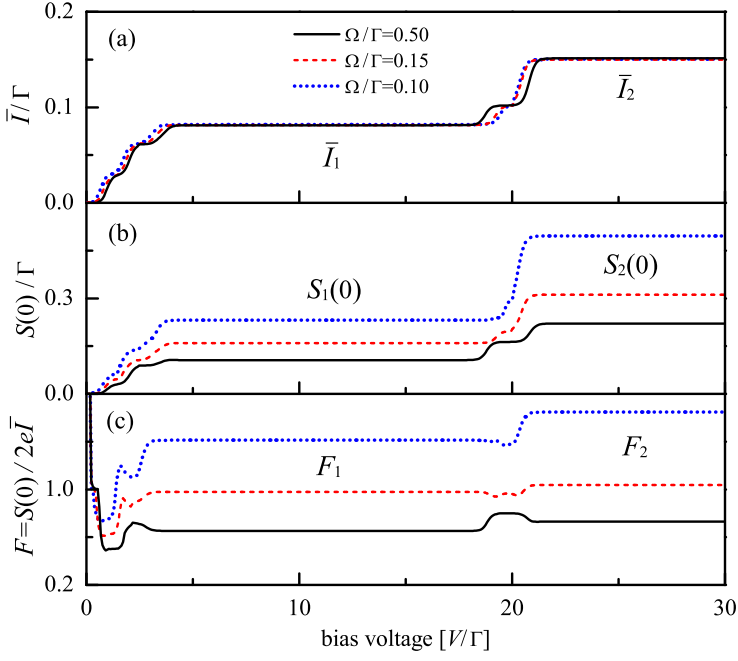}
\end{minipage}
\hspace{0.08\textwidth}
\begin{minipage}[c]{.38\textwidth}
\caption{\label{Fig2}(a) The measurement current $\bar{I}$,
 (b) zero frequency noise $S(0)$, and (c) Fano factor
 $F\equiv S(0)/(2e\bar{I})$
 versus the bias voltage for different values of $\Omg$.
 Each time when the Fermi level of the
 electrode aligns with one of the excitation energies as indicated
 in Table \ref{tab1}, a new transport channel opens.
 This leads to plateaus, separated by thermally broadened
 steps.
 The current and noise are measured in unit of $\Gamma\equiv\GamL+\GamR$.
 Other plotting parameters are: $\epl_\rmq=\epl_\rmT=0$,
 $W/\Gamma=0.5$, $k_{\rm B}T/\Gamma=2.0$, $\GamL/\GamR=1/3$,
 and $U/\Gamma=10$.}
\end{minipage}
\end{figure*}

 \section{\label{thsec3}Particle-Number-Resolved Quantum Master Equation Approach}

 The stochastic process of electron tunneling through the
 TDQD detector may be characterized by the joint
 probability distribution $P(\NL,\NR,t)$
 of finding $\NL$ electrons transmitted thought the left
 junction and $\NR$ electrons tunneled thought the right
 one in the given time $t$.
 Alternatively, it can be described by the current cumulants,
 known as full counting statistics \cite{Bla001,Naz03},
 which provides a unique signature of measurement characteristics.
 For that purpose, we employ a particle-number-resolved
 reduced density matrix $\rho^{(\NL,\NR)}$ for specific
 number of  $\NL(\NR)$ electrons passed through the left
 (right) junction.
 The corresponding particle-number-resolved QME reads
 \cite{Gur9615932,Agu04206601,Luo08345215,Li091707,Ema11085425,Fli08150601,Bra06026805}
 \begin{align}\label{CQME}
 \dot{\rho}^{(\NL,\NR)}\!=\!-\rmi{\cal L}\rho^{(\NL,\NR)}
 \!-\left\{{\cal R}_0\!+\!{\cal R}_\rmL\!+\!{\cal R}_\rmR
 \right\}\rho^{(\NL,\NR)},
 \end{align}
 where ${\cal L}(\cdots)\equiv[H_\rmS,(\cdots)]$ is the Liouvillian
 associated with the reduced system (qubit plus TDQD) Hamiltonian,
 \begin{widetext}
 \begin{subequations}
 \begin{gather}
 {\cal R}_0\rho^{(\NL,\NR)}=\frac{1}{2}\big\{|1\ra\la0|
 A^{(-)}\rho^{(\NL,\NR)}+\rho^{(\NL,\NR)}
 A^{(+)}|1\ra\la0|\big\}+{\rm h.c.}
 \end{gather}
 describes the continuous evolution of the reduced system,
 whereas
 \be
 {\cal R}_\rmL\rho^{(\NL,\NR)}=-\frac{1}{2}
 \big\{A^{(-)}_\rmL \rho^{(\NL-1,\NR)} |1\ra\la0|
 +|1\ra\la0| \rho^{(\NL+1,\NR)} A^{(+)}_\rmL\big\} +{\rm h.c.}
 \ee
 and
 \be
 {\cal R}_\rmR\rho^{(\NL,\NR)}=-\frac{1}{2}
 \big\{A^{(-)}_\rmR \rho^{(\NL,\NR-1)} |1\ra\la0|
 +|1\ra\la0| \rho^{(\NL,\NR+1)} A^{(+)}_\rmR\big\} +{\rm h.c.}
 \ee
 \end{subequations}
 \end{widetext}
 represent jumps of electrons via the left and right electrodes,
 respectively.
 Here $A^{(\pm)}=\sum_{\ell}A^{(\pm)}_\ell$, with
 $A^{(\pm)}_\ell\equiv C^{(\pm)}_\ell(\pm{\cal L})(|0\ra\la1|)$.
 The involving reservoir spectral functions are defined as the Fourier
 transform of the reservoir correlation functions
 \be
 C^{(\pm)}_\ell(\pm{\cal L})=\int_{-\infty}^\infty \rmd t
 C^{(\pm)}_\ell(t) e^{\pm \rmi {\cal L}t}.
 \ee

\begin{table*}
\caption{\label{tab2} The stationary current and Fano factor
 for $\epl_\rmq=\epl_\rmT=0$, and  $\Omg$, $W\ll U$
 in the bias regime 1: $ 2(E_4-E_2) > V > 2(E_5+E_2)$ and regime 2: $ V > 2(E_4+E_2) $.
 Owing to symmetric application of the bias voltage, the steps and hence
 the different bias regimes are distinguished at twice of the excitation
 energies as indicated in Table \ref{tab1}.}
\begin{ruledtabular}
\begin{tabular}{ccc}
 $i$ &    bias regime 1 & bias regime 2 \\ \hline
 $\bar{I}_i$&
 \parbox[c][1cm]{3.5cm}{\[\frac{\GamL\GamR}{3\GamL+2\GamR}\]}
 &
 \parbox[c][1cm]{3.5cm}{\[\frac{\GamL\GamR}{2\GamL+\GamR}\]}
 \\
 $F_i$ &
 \parbox[c][1cm]{8.6cm}
 {\[\frac{\GamL^2+4\GamR^2}{(3\GamL+2\GamR)^2}
 +\frac{\GamL^2\GamR^2(W^2+4\Omg^2)+4\GamL^2(W^4+4\Omg^4)}
 {2(3\GamL+2\GamR)^2\Omg^2W^2}\]}
 &
 \parbox[c][1cm]{8.6cm}
 {\[\frac{4\GamL^2+\GamR^2}{(2\GamL+\GamR)^2}
 +\frac{2\GamL^2(\GamR^2+8\Omg^2)\Omg^2+\GamL^2(\GamL^2+2W^2)W^2}
 {(2\GamL+\GamR)^2\Omg^2W^2}\]}
 \\
\end{tabular}
\end{ruledtabular}
\end{table*}

 The particle-number-resolved quantum master equation (\ref{CQME})
 provides us direct access to the joint probability
 distribution for the number of particles transmitted through
 the left and right junctions,
 i.e. $P(\NL,\NR,t)= {\rm tr} \{\rho^{(\NL,\NR)}(t)\}$, where
 ${\rm tr}\{\cdots\}$  represents the trace over the degrees of
 freedom of the reduced system (qubit plus TDQD).
 The first cumulant of the probability distribution corresponds 
 to the current through the left ($\ell=\rmL$) or right
 ($\ell=\rmR$) junction, given by
 $I_\ell=\frac{\rmd}{\rmd t}\sum_{\NL,\NR}N_\ell P(\NL,\NR)=
 {\rm tr}\{\frac{\rmd}{\rmd t}\hat{N}_\ell\}$,
 where
 $\hat{N}_\ell\equiv\sum_{\NL,\NR}N_\ell \rho^{(\NL,\NR)}$ can be
 evaluated via its equation of motion
 \bsube\label{Ntot}
 \be\label{Nt}
 \frac{\rmd}{\rmd t}\hat{N}_\ell=-\rmi{\cal L}\hat{N}_\ell
 -{\cal R}\hat{N}_\ell+{\cal T}^{(-)}_\ell \rho,
 \ee
 with
 \begin{gather}
 {\cal R}(\cdots)=\frac{1}{2}[|1\ra\la0|,A^{(-)}(\cdots)-(\cdots)
 A^{(+)}]+{\rm h.c.}
 \\
 {\cal T}_\ell^{(\pm)}(\cdots)=\frac{1}{2}[A^{(-)}_\ell(\cdots)
 |1\ra\la0| \pm |1\ra\la0|(\cdots)A^{(+)}_\ell]+{\rm h.c.}.
 \end{gather}
 \esube
 Straightforwardly, the measured current through junction
 $\ell$ is given by
 \be
 I_\ell(t)={\rm tr}\{{\cal T}_\ell^{(-)}\rho(t)\},
 \ee
 where $\rho(t)$ is the unconditional density matrix
 that satisfies
 \be\label{QME}
 \dot{\rho}=-\rmi{\cal L}\rho-{\cal R}\rho.
 \ee

 The second cumulant of the probability distribution
 is directly related to the shot noise.
 Here, we focus on the noise spectrum of circuit current.
 According to the Ramo-Shockley theorem \cite{Bla001}, the circuit current
 is given by $I(t)=\etaL I_{\rm L}+\etaR I_{\rm R}$.
 Here $\etaL$ and $\etaR$ are coefficients related to
 the junction capacitances that satisfy
 $\etaL+\etaR=1$  \cite{Bla001}.
 The transport currents through the left and right
 junctions are actually fluctuating in time, which give
 rise to charge accumulation ``$Q$'' on the TDQD.
 Due to charge conservation, it simply yields
 \be\label{Chconser}
 \dot{Q}=I_{\rm L}-I_{\rm R}.
 \ee
 One readily obtains the correlation function of circuit current
 \be
 I(t)I(0)=\etaL I_{\rm L}(t)I_{\rm L}(0)
 +\etaR I_{\rm R}(t)I_{\rm R}(0)-\etaL\etaR\dot{Q}(t)\dot{Q}(0).
 \ee
 As a result, the noise spectrum of circuit current consists of the following
 three parts \cite{Luo07085325,Gur05205341}
 \be\label{Stot}
 S(\omg)=\etaL S_{\rm L}(\omg)+ \etaR S_{\rm R}(\omg)-\etaL\etaR
 S_{\rm ch}(\omg),
 \ee
 where $S_{\rm L}(S_{\rm R})$  is the noise spectrum of the
 left (right) junction current, whereas $S_{\rm ch}(\omg)$
 stands for charge fluctuations in the TDQD.

 The noise spectrum of tunneling current $S_{\ell}(\omg)$
 ($\ell$=L or R) may be evaluated via the MacDonald's
 formula \cite{Mac62,Fli05411}
 \be\label{Salf}
 S_\ell(\omg)=\omega\int_0^\infty \rmd t \sin(\omg t)
 \frac{\rmd}{\rmd t}[\la \hat{N}_\ell^{2}(t)\ra-(\bar{I}t)^2].
 \ee
 Hereafter, it is assumed that the reduced system evolves from
 $t_0=-\infty$, such that reduced state at $t=0$, when
 measurement begins, have reached the stationary state
 $\rho_{\rm st}$. The involving current thus is a stationary
 one, i.e. $\bar{I}=I(t\rightarrow\infty)$.
 By employing the particle-number-resolved
 quantum master equation (\ref{CQME}), the quantity
 $\la\hat{N}_\ell^2(t)\ra\equiv{\rm tr}
 \{\sum_{\NL,\NR}N_\ell^2 \rho^{(\NL,\NR)}\}$
 is simply given by
 \bea\label{n2t}
 \frac{\rmd}{\rmd t}\la \hat{N}^2_\ell(t)\ra={\rm tr}\big\{2{\cal T}^{(-)}_\ell
 \hat{N}_\ell(t)+{\cal T}_\ell^{(+)}\rho_{\rm st}\big\},
 \eea
 where $\hat{N}_\ell(t)$ is obtained from \Eq{Ntot}.

 \begin{figure*}
 \includegraphics*[scale=1.2]{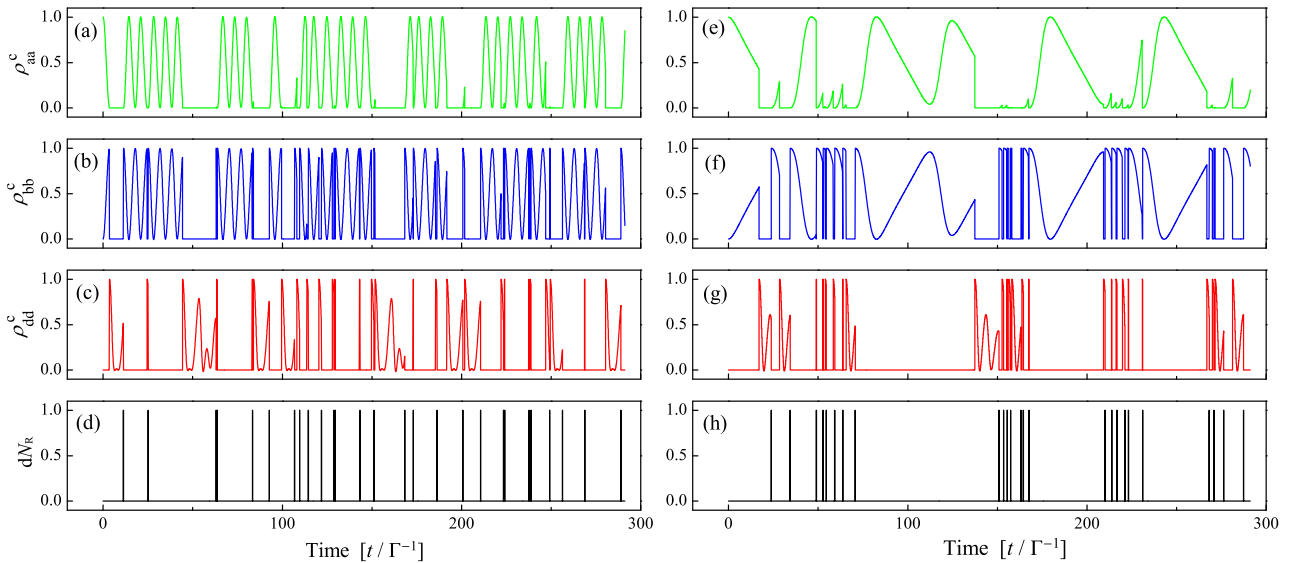}
 \caption{\label{Fig3} Sets of typical quantum trajectories and
 corresponding detection records for $\Omg/\Gam=0.5$ (a)-(d) and
 $\Omg/\Gam=0.1$ (e)-(h) with the same initial condition
 $\rho_{\rm aa}^\rmc(t=0)\equiv\la {\rm a}|\rho^\rmc(t=0)|{\rm a}\ra=1$.
 The measurement voltage $V/\Gam=10$ is within the bias regime 1
 [see \Fig{Fig2}], such that the state ``$|\rm c\ra$'' as shown
 in \Fig{Fig1}(c) is energetically prohibited.
 The Fermi energies are far from the excitation energies of the
 reduced system, such that the Fermi functions can be approximated
 by either 1 or 0, and thus temperature is not involved here.
 The time step used is $\Delta t$=0.01$\Gam^{-1}$.
 Other plotting parameters are: $\epl_\rmq=\epl_\rmT=0$,
 $W/\Gamma=0.5$, $\GamL/\GamR=1/3$,
 and $U/\Gamma=10$.}
 \end{figure*}

 For the charge fluctuations in the TDQD, the symmetrized
 spectrum reads \cite{Luo07085325}
 \be\label{Sch}
 S_{\rm ch}(\omg)=\omg^2\int_{-\infty}^{\infty}\rmd\tau\la Q(\tau)Q
 +QQ(\tau)\ra e^{\rmi\omg\tau},
 \ee
 where $Q\equiv|1\ra\la1|+|2\ra\la2|$ stands for the operator of electron charge
 on the TDQD, and $\la Q(\tau)Q\ra\equiv{\rm tr}\{{\rm tr}_{\rm B}[U^\dag(\tau)Q
 U(\tau)Q\rho_{\rm st}\rho_{\rm B}]\}$, with $U(\tau)$ being the evolution
 operator associated with the entire system Hamiltonian (\ref{Htot}).
 By introducing an alternative reduced density matrix
 $\tilde{\rho}(\tau)\equiv{\rm tr}_{\rm B}[U(\tau)Q\rho_{\rm st}\rho_{\rm B}
 U^\dag(\tau)]$, the charge correlation can be further reduced to
 $\la Q(\tau)Q\ra={\rm tr}\{Q\tilde{\rho}(\tau)\}$.
 Under the second-order Born-Markov approximation, it is found that
 $\tilde{\rho}(t)$ satisfies the same equation as $\rho(t)$ in \Eq{QME},
 with the only crucial difference of the initial condition
 $\tilde{\rho}(0)=Q\rho_{\rm st}$.
 Eventually, the noise spectrum of charge fluctuations reads
 \be
 S_{\rm ch}(\omg)=2\omg^2{\rm Re}\{{\rm tr}[
 Q\tilde{\rho}(\omg)+Q\tilde{\rho}(-\omg)]\},
 \ee
 where $\tilde{\rho}(\omg)$ is the Fourier transform of
 $\tilde{\rho}(t)$ and satisfies
 \be
 -\rmi\omg \tilde{\rho}(\omg)=-\rmi{\cal L}\tilde{\rho}(\omg)
 -{\cal R}\tilde{\rho}(\omg)+Q\rho_{\rm st}.
 \ee.

 \section{\label{thsec4}Qubit Dynamics Induced Super-Poissonian Noise}

 The measurement current $\bar{I}$, zero frequency noise $S(0)$,
 and the Fano factor $F=S(0)/(2e\bar{I})$ versus voltage are
 plotted in \Fig{Fig2}(a)-(c), respectively.
 At very low bias $V\ll k_{\rm B}T$, electron transport through
 the TDQD detector is exponentially suppressed. The current
 fluctuation is dominated by thermal noise described by the
 hyperbolic cotangent behavior \cite{Bla001},
 which leads to a divergence of the Fano factor at $V=0$, as
 shown in \Fig{Fig2}(c).
 Each time when a new excitation energy (as indicated in Table \ref{tab1})
 lies within the energy window defined by the chemical potentials of the
 left and right electrodes, a new transport channel opens, which gives
 rise to plateaus, separated by thermally broadened steps.
 Owing to symmetric application of the bias, the steps take place
 at bias voltages twice of the corresponding excitation energy.

 The plateau heights of the current are found to be
 independent on $\Omg$.
 Variation of $\Omg$ changes the eigenenergies (see Table \ref{tab1}),
 leading thus only to small shift of the current steps, as
 displayed in \Fig{Fig2}(a).
 The plateau heights of noise and Fano factor, however, are
 sensitively modulated by $\Omg$, showing shot noise as a
 more sensitive diagnostic tool than the current.
 For $\Omg/\Gamma=0.5$, the noise is well below the Poissonian
 value.
 An decrease in $\Omg$ leads to a strong enhancement of the Fano
 factor. In particular, a prominent super-Poissonian noise is
 observed for $\Omg/\Gam=0.1$, as shown by the dotted curve
 in \Fig{Fig2}(c).
 In literature, different mechanisms responsible for super-Poissonian
 noise have been identified, such as
 dynamical channel blockade \cite{Urb09165319,Lu10034314,Luo11145301,Wan11115304},
 dynamical spin blockade \cite{Wu10125326,Luo14892,Ubb13041304},
 or cotunneling events \cite{Mis09224420,Wey11195302,Oka13041302}.
 Our result reveals that the intrinsic dynamics of the qubit
 serves as an additional mechanism that may lead to
 super-Poissonian shot noise in a double dot detector.

 Specifically, let us investigate the current and noise in the
 bias regime 1: $ 2(E_4-E_2) > V > 2(E_5+E_2)$ and regime 2:
 $ V > 2(E_4+E_2) $.
 Here the factor of ``2'' arises from the symmetric application
 of the bias voltage.
 In these two wide regions where electrode chemical potentials
 are far away from the excitation energies of the double dot,
 the Fermi functions can be well approximated by either one or zero.
 Analytical results of the current and noise are obtained for
 $\epl_\rmq=\epl_\rmT=0$, and $\Omg$,  $W\ll U$,
 as listed in Table \ref{tab2}.
 Indeed, the current plateau height
 depend on the coupling parameters $\GamL$ and $\GamR$ only,
 while noise and Fano factor are both sensitive to the
 interdot couplings $\Omg$ and $W$.
 Strikingly, a divergent Fano factor is found in the limit
 $\Omg\rightarrow0$ or $W\rightarrow0$.

 \begin{figure*}
 \includegraphics*[scale=0.9]{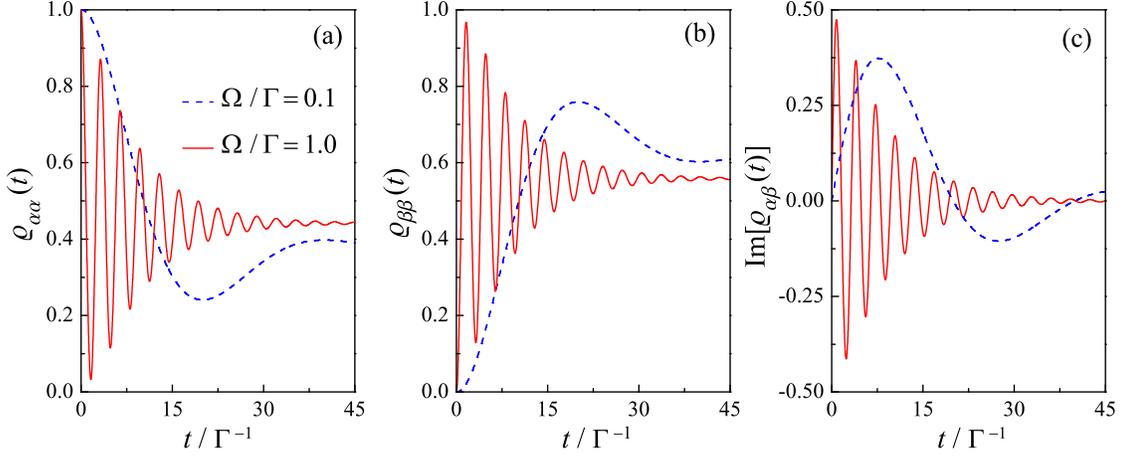}
 \caption{\label{Fig4}
 Measurement-induced qubit relaxation and dephasing versus time
 for $\Omg/\Gam=0.1$ (dashed curves) and $\Omg/\Gam=1.0$ (solid curves).
 (a) $\varrho_{\alpha\alpha}(t)$, (b) $\varrho_{\beta\beta}(t)$,
 (c) the imaginary part of the off-diagonal element
 $\varrho_{\alpha\beta}$.
 The total tunneling width $\Gam=\GamL+\GamR$ is kept constant,
 and interdot hopping in the TDQD is $W/\Gam=1.0$.
 The qubit is assumed to be
 symmetric ($\epl_\rmq=0$) and initially in the logical
 state $|\alpha\ra$, i.e. $\varrho(t=0)=|\alpha\ra\la\alpha|$.
 Other plotting parameters used are: $V/\Gam=10$, $\epl_\rmT=0$,
 $\GamL/\GamR=1/4$ and $U/\Gamma=10$.}
 \end{figure*}

 To investigate in detail the underlying physics that leads to
 the divergent Fano factor, we now resort to the
 real-time measurement dynamics of the reduced system, i.e.,
 the usual single measurement realizations in
 experiments.
 In what follows, we will consider a typical voltage $V/\Gam=10$ in the bias regime 1
 as shown in \Fig{Fig2}, such that the state ``$|\rmc\ra$'' [\Fig{Fig1}(c)]
 is energetically prohibited.
 The reason we consider this regime is that the measured current
 visibility, defined as $|I_\alf-I_\beta|/(I_\alf+I_\beta)$, can
 reach the maximum value of 1. Here $I_\alf$ ($I_\beta$) stands
 for the current through the TDQD when the qubit occupies the
 logical state $|\alf\ra$ ($|\beta\ra$).
 Electrons flow in one direction: An extra
 electron injects into the QD1 from the left electrode, dwells in
 the double dot for a certain amount of time before it escapes to the
 right electrode.
 We introduce two stochastic point variables d$N_{\rmL}(t)$ and
 d$N_{\rmR}(t)$ (with values either 0 or 1)
 to represent, respectively, the number of electron tunneled
 into QD1 from the left electrode and that escaped
 to the right electrode from the QD1, during the infinitesimal
 time interval d$t$.
 According to the quantum trajectory theory, the evolution of
 the reduced system is given by the
 following conditional QME \cite{Goa01125326}
 \begin{widetext}
 \begin{align}\label{rhoc}
 \rmd \rho^\rmc=&-\rmi {\cal L}\rho^\rmc(t)\rmd t
 -\{\Gam_{\rmL}{\cal A}[|1\ra\la0|]+\Gam_{\rmR}{\cal A}[|0\ra\la1|]
 -{\cal P}_{\rmL}(t)-{\cal P}_{\rmR}(t)\}\rho^\rmc(t)\rmd t
 \nl
 &
 +\rmd N_{\rmL}\left[\frac{{\cal J}[\sqrt{\GamL}|1\ra\la0|]}
 {{\cal P}_{\rmL}(t)}-1\right]\rho^{\rmc}(t)
 +\rmd N_{\rmR}\left[\frac{{\cal J}[\sqrt{\GamR}|0\ra\la1|]}
 {{\cal P}_{\rmR}(t)}-1\right]\rho^{\rmc}(t),
 \end{align}
 \end{widetext}
 where we have introduced the superoperators
 ${\cal J}[X]\rho^{\rmc}\equiv X\rho^{\rmc}X^\dag$ and
 ${\cal A}[X]\rho^{\rmc}\equiv\frac{1}{2}(X^\dag X\rho^{\rmc}+\rho^{\rmc}X^\dag X)$.
 The attached superscript ``c'' to the reduced density matrix
 is to specify that its evolution is conditioned on the measurement results.
 A simple ensemble average over a large
 number of particular realizations of $\rho^\rmc(t)$ would recover the
 unconditional density matrix $\rho(t)$ in \Eq{QME}, i.e. $\rho(t)=E[\rho^\rmc(t)]$,
 where $E[X]$ stands for an ensemble average of a large number
 of quantum trajectories.
 The involving stochastic variables for single electron tunneling
 events satisfy
 \bsube\label{dN}
 \begin{gather}
 E[\rmd N_{\rmL}(t)]={\cal P}_{\rmL}(t)\rmd t
 =\Tr\{ {\cal J}[\sqrt{\Gam_{\rmL}}|1\ra\la0|]\rho^{\rmc}\}\rmd t,\label{dNL}
 \\
 E[\rmd N_{\rmR}(t)]={\cal P}_{\rmR}(t)\rmd t
 =\Tr\{ {\cal J}[\sqrt{\Gam_{\rmR}}|0\ra\la1|]\rho^{\rmc}\}\rmd t.\label{dNR}
 \end{gather}
 \esube
 It is now clear that individual electron tunneling events
 condition the future evolution of the reduced density
 matrix [\Eq{rhoc}], while instantaneous quantum state
 conditions the detected tunneling events through the left
 and right junctions [\Eq{dN}].
 By employing this approach, one thus is capable of
 propagating the conditioned quantum state [$\rho^\rmc(t)$]
 and measurement result [d$N_{\rmL/\rmR}(t)$] in a
 self-consistent way.

 The real-time quantum state [$\rho^\rmc(t)$] and
 corresponding detection record of tunneling to the right
 electrode [${\rm d}N_{\rm R}(t)$] are plotted in
 \Fig{Fig3}(a)-(d) for $\Omg/\Gam=0.5$.
 For a give voltage $V/\Gam=10$ in the bias regime 1, the state ``$|\rm c\ra$''
 as shown in \Fig{Fig1}(c) is energetically forbidden.
 When there is no extra electron in the TDQD, the qubit
 experiences some oscillations between the states ``$|\rm a\ra$''
 and ``$|\rm b\ra$'' shown in \Fig{Fig1} with frequency $\sim\Omg$.
 Whenever one electron tunnels into the TDQD, the system collapses
 to the state ``$|\rm d\ra$''.
 The electron may stay in the double dot and experience some
 oscillations between QD1 and QD2, until it escapes to the
 right electrode.
 Correspondingly, the system jumps to the state ``$|\rm b\ra$'',
 and an event of tunneling out to the right electrode is detected,
 i.e. ${\rm d}N_{\rm R}=1$.
 A typical example of the tunneling events is shown in \Fig{Fig3}(d).

 Very different tunneling behavior is observed in the case of
 a suppressed $\Omg$; see \Fig{Fig3}(e)-(h) for $\Omg/\Gam=0.1$.
 One finds unambiguously the bunching of electron tunneling events
 though the TDQD.
 In most of the time, the system is oscillating between the
 states ``$|\rm a\ra$'' and ``$|\rm b\ra$'' with a lower
 frequency $\sim \Omg$.
 Due to strong electrostatic interaction between the qubit
 and TDQD ($U/\Gam\gg 1$),
 the occupation of qubit in the logical state ``$|\alf\ra$''
 blocks the current through TDQD until it tunnels to the
 state ``$|\beta\ra$'', which is then followed by a bunching
 of tunneling events through the TDQD during a short time
 window.
 It is right this mechanism that leads to the super-Poissonian
 Fano factor in \Fig{Fig2}.
 Our result thus reveals that the intrinsic dynamics of the qubit
 may serves as an additional mechanism that may lead to
 dynamical blockade and eventually to the pronounced
 super-Poissonian behavior in noise spectrum.

 Normally, detector shot noise leads to the dephasing
 of a qubit \cite{Gur05073303}.
 However, we will show in \sref{thsec5} that a large detector shot noise
 at small $\Omg$ does not necessarily imply a fast dephasing rate.
 In particular, it will be revealed that the dynamical blockade may
  have essential roles to play in the dephasing process of the qubit.
 The dephasing is suppressed at small $\Omg$, in spite of a large
 detector shot noise.

 \section{\label{thsec5}Detection Backaction Induced Dephasing}

 To study the dephasing behavior of the qubit under continuous
 measurement of the TDQD detector, we shall make
 use of the density matrix of the qubit alone, which can be obtained
 by tracing out the degrees of freedom of the
 TDQD from the reduced (qubit plus TQDQ) density matrix
 \be\label{varrho}
 \varrho(t)={\rm tr}_{\rm TDQD}\{\rho(t)\},
 \ee
 where ${\rm tr}_{\rm TDQD}\{\cdots\}$ stands for the trace over the
 degrees of freedom of the TDQD, and $\rho(t)$ is the
 unconditional density matrix that can be obtained from \Eq{QME}.
 To obtain time evolution of qubit state alone $\varrho(t)$, it is thus necessary to
 derive the equation of motion $\rho(t)$ first.
 In the state representation of \Fig{Fig1},
 the quantum master equation of $\rho(t)$ is given by
 \begin{widetext}
 \bsube\label{EOM-rho}
 \begin{gather}
 \dot{\rho}_{\rm aa}=\rmi \Omg (\rho_{\rm ab}-\rho_{\rm ba})
 +\GamL\rho_{\rm cc}+\GamR\rho_{\rm cc},
\\
 \dot{\rho}_{\rm bb}=\rmi \Omg (\rho_{\rm ba}-\rho_{\rm ab})
 -\GamL\rho_{\rm bb}+\GamR\rho_{\rm dd},
\\
 \dot{\rho}_{\rm cc}=\rmi \Omg (\rho_{\rm cd}-\rho_{\rm dc})
 -\GamL\rho_{\rm cc}-\GamR\rho_{\rm cc}+\rmi W (\rho_{\rm ce}-\rho_{\rm ec}),
\\
 \dot{\rho}_{\rm dd}\!=\rmi \Omg (\rho_{\rm dc}-\rho_{\rm cd})
 \!+\GamL\rho_{\rm bb}\!-\GamR\rho_{\rm dd}+\rmi W (\rho_{\rm df}-\rho_{\rm fd}),
\\
 \dot{\rho}_{\rm ee}= \rmi\Omg (\rho_{\rm ef}-\rho_{\rm fe})
 -\rmi W(\rho_{\rm ce}-\rho_{\rm ec}),
\\
 \dot{\rho}_{\rm ff}= \rmi\Omg (\rho_{\rm fe}-\rho_{\rm ef})
 -\rmi W (\rho_{\rm df}-\rho_{\rm fd}),
\\
 \dot{\rho}_{\rm ab}=\rmi \Omg (\rho_{\rm aa}-\rho_{\rm bb})
 -{\textstyle \frac{1}{2}}\GamL(\rho_{\rm ab}-\rho_{\rm cd})+\GamR\rho_{\rm cd},
\\
 \dot{\rho}_{\rm cd}=\rmi \Omg (\rho_{\rm cc}-\rho_{\rm dd})
 +\rmi W (\rho_{\rm cf}-\rho_{\rm ed})
 -\rmi U \rho_{\rm cd}+{\textstyle \frac{1}{2}}\GamL(\rho_{\rm ab}-\rho_{\rm cd})
 -\GamR\rho_{\rm cd},
\\
 \dot{\rho}_{\rm ce}=\rmi\Omg (\rho_{\rm cf}-\rho_{\rm ed})
 +\rmi W (\rho_{\rm cc}-\rho_{\rm ee})
 -\rmi U \rho_{\rm ce}-{\textstyle \frac{1}{2}}(\GamL+\GamR)\rho_{\rm ce},
\\
 \dot{\rho}_{\rm cf}=\rmi\Omg (\rho_{\rm ce}-\rho_{\rm df})
 +\rmi W (\rho_{\rm cd}-\rho_{\rm ef})
 -\rmi U \rho_{\rm cf}-{\textstyle \frac{1}{2}}(\GamL+\GamR)\rho_{\rm cf},
\\
 \dot{\rho}_{\rm de}=\rmi\Omg (\rho_{\rm df}-\rho_{\rm ce})
 +\rmi W (\rho_{\rm dc}-\rho_{\rm fe})-{\textstyle \frac{1}{2}}\GamR\rho_{\rm de},
\\
 \dot{\rho}_{\rm df}=\rmi\Omg (\rho_{\rm de}-\rho_{\rm cf})
 +\rmi W (\rho_{\rm dd}-\rho_{\rm ff})-{\textstyle \frac{1}{2}}\GamR\rho_{\rm df},
\\
 \dot{\rho}_{\rm ef}=\rmi\Omg (\rho_{\rm ee}-\rho_{\rm ff})
 -\rmi W (\rho_{\rm cf}-\rho_{\rm ed}).
 \end{gather}
 \esube
 \end{widetext}
 From the above coupled equations, one then is able to obtain the
 reduced dynamics of the qubit alone by using \Eq{varrho}, i.e. 
 $\varrho_{\alf\alf}=\rho_{\rm aa}+\rho_{\rm cc}+\rho_{\rm ee}$,
 $\varrho_{\beta\beta}=\rho_{\rm bb}+\rho_{\rm dd}+\rho_{\rm ff}$,
 and $\varrho_{\alf\beta}=\rho_{\rm ab}+\rho_{\rm cd}+\rho_{\rm ef}$,
 representing the probability of the qubit in the
 logical states $|\alf\ra$, $|\beta\ra$, and linear superposition
 of the two logical states (so-called ``quantum coherence''), respectively.
 By collecting relevant terms in \Eq{EOM-rho}, one eventually arrives at the
 equation of motion for the reduced density matrix of the qubit
 \bsube\label{varrhotot}
 \begin{gather}
 \dot{\varrho}_{\alf\alf}(t)=\rmi \Omg (\varrho_{\alf\beta}-\varrho_{\beta\alf}),
 \label{varrhoaa}
 \\
 \dot{\varrho}_{\beta\beta}(t)=\rmi \Omg (\varrho_{\beta\alf}-\varrho_{\alf\beta}),
 \label{varrhobb}
 \\
 \dot{\varrho}_{\alf\beta}(t)=\rmi \Omg (\varrho_{\alf\alf}-\varrho_{\beta\beta})
 -\rmi U \rho_{{\rm c d}}. \label{varrhoab}
 \end{gather}
 \esube
 \Eqs{varrhoaa} and (\ref{varrhobb}) denote coherent oscillations of
 the qubit, while \Eq{varrhoab} stands for the dephasing of the qubit.
 Unambiguously, the qubit dephasing is directly related to $\rho_{\rm cd}$
 [cf. the last term in \Eq{varrhoab}], which is further coupled to the
 dynamics of the entire system as shown in \Eq{EOM-rho}.
 It thus implies that the dynamics of the qubit and that of the TDQD
 are intimately entangled.
 Physically, due to detector current transport through two discrete
 levels of the TDQD, an electron tunneled into the TDQD is a linear
 superposition of these two states;
 the qubit itself is a two-state system described by superposition,
 leading thus eventually to the entanglement between the qubit and
 TDQD.

 \begin{figure*}
 \includegraphics*[scale=0.85]{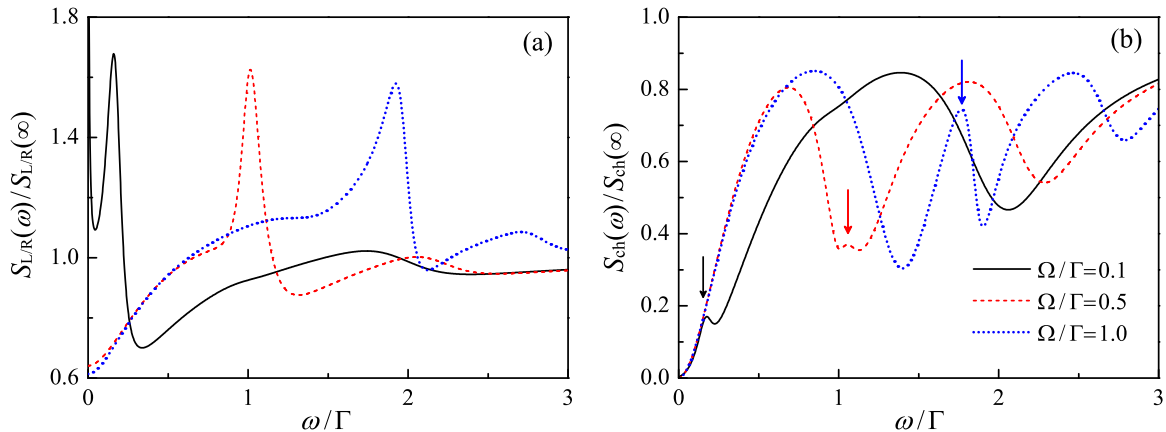}
 \caption{\label{Fig5}
 (a) Noise spectrum of the tunneling current through the left or
 right electrodes $S_{\rm L/R}(\omega)$, scaled by its own pedestal
 $S_{\rm L/R}(\infty)$.
 (b) Spectrum of charge fluctuation $S_{\rm ch}(\omega)$ in the TDQD
 with respective to its pedestal $S_{\rm ch}(\infty)$.
 The total tunneling width $\Gam=\GamL+\GamR$ is kept constant.
 Other parameters are: $\epl_\rmq=\epl_\rmT=0$, $W/\Gam=1.0$,
 $V/\Gamma=10$, and $U/\Gamma=10$.}
 \end{figure*}

 It was revealed that for a DQD detector, the qubit dephasing rate
 is directly related to the strength of the coupling between DQD 
 and the left or right electrodes ($\GamL$ or $\GamR$), rather than
 the interdot coupling of the qubit ($\Omg$) \cite{Jia07155333}.
 We will show, however, the interdot coupling of the qubit may also 
 have essential roles to play in the dephasing process of the qubit
 itself.
 It is thus instructive to study qubit dynamics at different values 
 of $\Omg$.
 The numerical results, obtained by propagating \Eqs{EOM-rho} and
 (\ref{varrhotot}) in parallel, are displayed in \Fig{Fig4}.
 Coherent oscillations of the qubit are shown in \Fig{Fig4}(a)
 $\varrho_{\alpha\alpha}(t)$ and (b) $\varrho_{\beta\beta}(t)$,
 respectively.
 The dephasing of the qubit, described by the off-diagonal
 element of the reduced density matrix $\varrho_{\alf\beta}(t)$,
 is plotted in \Fig{Fig4}(c) for $\Omg/\Gam=0.1$ (dashed curve)
 and $\Omg/\Gam=1.0$ (solid curve).
 In both cases, $\varrho_{\alf\beta}(t)$ vanishes in the long time limit,
 leading thus to the ``collapse'' of the reduced density matrix
 into the statistical mixture.
 However, the dephasing processes are indeed very different for the
 two cases.
 It is found via numerical fitting that the dephasing rate for
 $\Omg/\Gam=1.0$ could reach almost 4 times larger that that for
 $\Omg/\Gam=0.1$.
 Our result thus shows qubit interdot coupling ($\Omg$) as an essential
 mechanism that may influence qubit dephasing, complementary to the
 conventional ways in SET measurement.

 The unique suppression of the dephasing at small $\Omg$
 can be interpreted as follows.
 In the case of a large interdot coupling $\Omg$ (for instance, $\Omg/\Gam=1.0$),
 electrons tunnel through the TDQD very frequently; see individual electron
 tunneling events in \Fig{Fig2} (d).
 It thus gives rise to a frequent perturbation (measurement) of the qubit.
 In contrast, for a small $\Omg$ (cf. $\Omg/\Gam=0.1$), electron transport
 through the TDQD is dynamically blockaded during the time windows where coherent
 oscillations between the states ``$|{\rm a}\ra$'' and ``$|{\rm b}\ra$'' dominates,
 as shown in \Fig{Fig2}(e)-(h).
 Yet, these time windows corresponds to no-measurement regimes
 where acquisition of qubit information is suppressed, leading eventually
 to the inhibition of the dephasing.

 As is well known, the fundamental physics involved in quantum
 detection is the trade-off between acquisition of qubit information
 and the backaction-induced dephasing of the measured system.
 A question arises naturally for the present TDQD detector:
 Is the measurement more effective
 in the small $\Omg$ regime where dephasing is suppressed, or in the
 large $\Omg$ regime where the measurement takes place more frequently?
 Thus, we now investigate the effectiveness of the measurement at
 different values of qubit interdot coupling in \sref{thsec6}.

\section{\label{thsec6}Measurement Effectiveness}

 A powerful tool to characterize the measurement effectiveness
 is the detector's noise spectrum $S(\omg)$.
 The qubit oscillations are manifested in $S(\omg)$ as a peak
 located at the qubit characteristic frequency
 $\omg_\rmc=(\epl_\rmq^2+4\Omg^2)^{1/2}$.
 An essential feature of this peak is that its height with
 respective to the pedestal, also known as  signal-to-noise ratio,
 provides a measure of detector's effectiveness, showing how close
 to the quantum limit the detector may operate \cite{Kor01085312,Kor01165310}.
 It was argued that for any linear-response detectors there is a 
 fundamental limit imposed on the signal-to-noise ratio, i.e. the 
 so-called Korotkov-Averin bound \cite{Kor01085312,Kor01165310}: 
 The peak height can reach maximally
 4 times the noise pedestal for an ideal or quantum-limited detector.
 In contrast, for a less efficient detector the qubit dephasing
 takes place more rapidly than information acquisition, and the
 signal-to-noise ratio is less than four.

 To analyze the signal-to-noise ratio of a TDQD detector, we first study
 the noise spectrum of the tunneling currents through the
 left or right junction $S_{\rmL/\rmR}(\omg)$ [cf. \Eq{Salf}].
 The numerical result is displayed in \Fig{Fig5}(a), where the noise
 of the tunneling current is scaled by its own pedestal
 \be\label{SLRP}
 S_{\rm L/R}(\infty)=\frac{2\GamL\GamR}{3\GamL+2\GamR}.
 \ee
 The noises of left and right junction currents are found to
 be consistent within the whole frequency regime, i.e.
 $S_\rmL(\omg)=S_\rmR(\omg)$.
 The noise at various values of interdot couplings ($\Omg$)
 is plotted in \Fig{Fig5}(a).
 The peaks in vicinity of $\omg\approx2\Omg$ reflect signal
 of qubit coherent oscillations.
 The peak width increases with rising $\Omg$, indicating
 the enhancement of the dephasing rate.
 It thus confirms our previous argument of the dependence
 of the dephasing on qubit interdot coupling.

 What we are most interested is the height of the peak of qubit
 oscillations, which provides the measure of signal-to-noise
 ratio of quantum measurement.
 For the present TDQD detector, it is found in \Fig{Fig5} that the
 peak height at different values of $\Omg$ does not show striking
 difference.
 Although $\Omg$ has essential roles to play in the dephasing of the
 qubit, its influence on signal-to-noise ratio is very limited.
 At small $\Omg$, dephasing is inhibited but information acquisition
 is also suppressed.
 An increase of $\Omg$ leads to fast information gain, whereas the
 qubit lose coherence more rapidly.
 Eventually, the measurement effectiveness turns out to be insensitive
 to the qubit interdot coupling ($\Omg$).
 Furthermore, by considering the noise spectrum of the tunneling currents,
 the signal-to-noise is found to be well below the Korotkov-Averin
 bound.
 This is qualitatively consistent with the result in
 Ref. \onlinecite{Gil06116806}, where the signal-to-noise ratio
 of a serial DQD detector is found below ``4''.
 It might lead us to conclude that neither the TDQD nor the serial DQD 
 can reach the effectiveness of an ideal detector, if one takes solely 
 the tunneling current noise into consideration.

 However, this picture is not yet complete for a TDQD detector, since
 the currents through the left and right junctions are intrinsically
 correlated via the charge accumulation in the TDQD, owing to the
 condition of charge conservation [cf. \Eq{Chconser}].
 The noise of the circuit current is actually a superposition of
 each component; see \Eq{Stot}.
 It is thus of importance to study the influence of charge
 fluctuation [$S_{\rm ch}(\omg)$] on the signal-to-noise ratio.
 In particular, we will show in \sref{thsec7} that under appropriate
 conditions the charge fluctuation leads to a strong suppression of
 the noise pedestal. It gives rise to a strong enhancement
 of the signal-to-noise ratio, leading eventually to the violation
 of the Korotkov-Averin bound.

\section{\label{thsec7}Violation of the Korotkov-Averin Bound}

 \Fig{Fig5}(b) shows the numerical result of the charge
 fluctuations in the TDQD for various values of $\Omg$.
 The plot of the charge fluctuation is scaled by its own
 pedestal
 \be\label{SchP}
 S_{\rm ch}(\infty)=\frac{8\GamL\GamR}{3\GamL+2\GamR}.
 \ee
 In the low frequency limit, the charge fluctuation
 in the TDQD is strongly inhibited, as implied in \Eq{Sch}.
 The basic signals are the peaks located in the vicinity
 of frequency $2\Omg$, indicating qubit coherent oscillations;
 see the arrows in \Fig{Fig5}(b).

\begin{figure}
\includegraphics[scale=1.15]{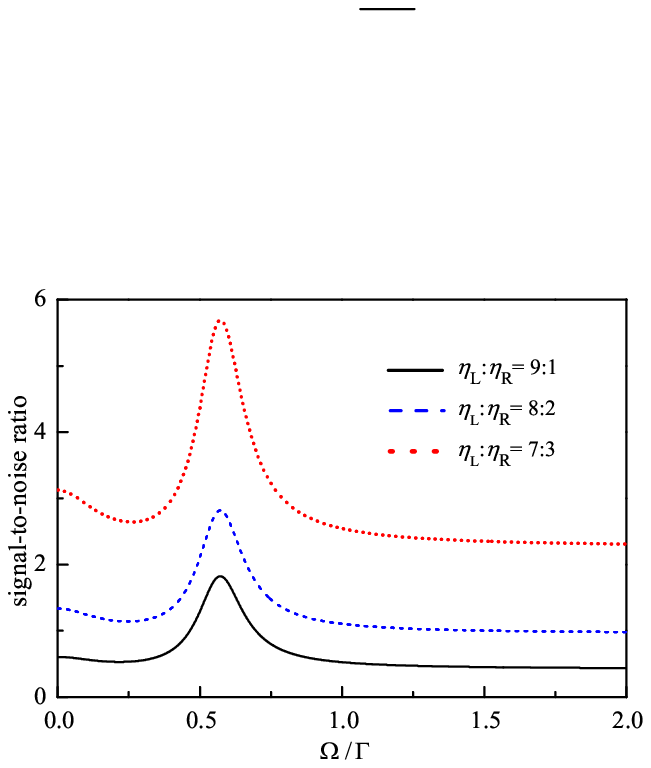}
\caption{\label{Fig6}
 The signal-to-noise ratio versus $\Omg$ obtained from the circuit current
 noise with different configurations of left and right junction capacitances
 ($\eta_{\rm L}:\eta_{\rm R}$).
 The coefficients satisfies $\eta_{\rm L}+\eta_{\rm R}=1$.
 The total tunneling width $\Gam=\GamL+\GamR$ is kept constant.
 Other plotting parameters used are: $\epl_\rmq=\epl_\rmT=0$, $W/\Gam=1.0$,
 $V/\Gamma=10$, $\GamL/\GamR=0.5$, and $U/\Gamma=10$.}
\end{figure}

 The charge fluctuation may have a significant impact on the
 signal-to-noise ratio of the measurement, as displayed in \Fig{Fig6}.
 In case of very asymmetric junction capacitances, for instance,
 $\eta_\rmL:\eta_\rmR=9:1$ or $\eta_\rmL:\eta_\rmR=8:2$,
 the spectrum of charge fluctuations has only very limited
 contribution to the circuit noise [cf. \Eq{Stot}]. The resultant
 signal-to-noise ratio is below the Korotkov-Averin bound;
 see the solid and dashed curves in \Fig{Fig6}.
 However, when the junction capacitances get more and more
 symmetric, the charge fluctuation may have a vital role to play in
 the noise of circuit current.
 Strikingly, for $\eta_\rmL:\eta_\rmR=7:3$, a prominent enhancement
 of the signal-to-noise ratio is observed and the signal-to-noise
 ratio exceeds the upper limit of ``4'',
 i.e. the violation of the Korotkov-Averin bound.
 Furthermore, by checking the pedestals of the tunneling current
 noise [\Eq{SLRP}] and charge fluctuation [\Eq{SchP}], one finds
 that for symmetric junction capacitances ($\eta_\rmL:\eta_\rmR=1:1$),
 the pedestal of the circuit noise [\Eq{Stot}] can be completely
 eliminated, resulting thus in a divergence of the signal-to-noise ratio.
 Our new finding thus provides a transparent and direct way to improve
 the signal-to-noise ratio of a TDQD detector.

 In literature, different approaches have been proposed that may lead
 to the violation of the Korotkov-Averin bound. Normally, they fall into two
 categories. The first type is based on the enhancement of measurement
 signal by employing approaches such as quantum nondemolition
 measurements \cite{Ave02207901,Jor05125333}, non-Markovian memory
 effect \cite{Luo13173703}, or quantum feedback
 scheme \cite{Wan07155304,Vij1277}.
 The second on concerns with the reduction of the noise pedestal
 by utilizing twin detectors \cite{Jor05220401,Kun09035314}, or strongly responding
 detectors \cite{Jia09075320}.
 The occurrence of a divergent signal-to-noise ratio in this work arises from a
 complete suppression of the noise pedestal.
 Our result shows that by simply adjusting the junction capacitances,
 one may considerably enhance the signal-to-noise ratio of a TDQD
 detector in quantum measurement.

\section{\label{thsec8}Summary}

 We have proposed to continuously monitor a charge qubit by
 utilizing a T-shaped double quantum dot detector, in which
 only one dot is directly tunnel-coupled to the electrodes.
 It is demonstrated that the dynamics of the qubit and the detector
 output are intrinsically correlated.
 In case of a suppressed interdot coupling between the two
 states of the qubit, a dynamical blockade mechanism takes
 place, leading to a super-Poisson shot noise.
 However, such a pronounced enhancement of the noise does not
 necessarily produce a fast dephasing rate.
 Actually, an inhibited dephasing is observed, since the involving
 dynamical blockade is directly related to the regime where no
 information is acquired.
 The major advantage of the present T-shaped double quantum dot
 detector is that its spectrum of charge fluctuations
 may significantly suppress the pedestal of the circuit noise.
 Remarkably, the noise pedestal could be removed completely
 under the condition of symmetric junction capacitances, leading
 to a divergent signal-to-noise ratio, and eventually to the violation
 the Korotkov-Averin bound in quantum measurement.
 The proposed TDQD thus may serve as an essential candidate
 detector to reach the quantum limited effectiveness in a very
 transparent and straightforward manner.

 \begin{acknowledgments}
 Support from the National Natural Science Foundation of
 China (11147114 and 11204272) and the Natural Science Foundation of
 Zhejiang Province (Y6110467)
 are gratefully acknowledged.
 \end{acknowledgments}


\end{document}